\documentclass[journal]{IEEEtranTIE}
\usepackage{graphicx}
\usepackage{picinpar}
\usepackage{amsmath}
\usepackage{url}
\usepackage{flushend}
\usepackage{colortbl}
\usepackage{soul}
\usepackage{multirow}
\usepackage{pifont}
\usepackage{color}
\usepackage{alltt}
\usepackage[hidelinks]{hyperref}
\usepackage{enumerate}
\usepackage{siunitx}
\usepackage{epstopdf}
\usepackage{pbox}
\usepackage{amsmath,amssymb,amsfonts}

\usepackage{algorithmic}
\usepackage{graphicx}
\usepackage{textcomp}
\usepackage{xcolor}
\usepackage{cuted}
\usepackage{stfloats}
\usepackage[RPvoltages]{circuitikz}
\usepackage[thinc]{esdiff}
\usepackage{tikz}
\usepackage{import}
\usetikzlibrary{arrows,matrix,positioning,fit}
\usetikzlibrary{calc}
\usetikzlibrary{shapes}
\ctikzset{bipoles/cuteinductor/coils=10}
\ctikzset{bipoles/cuteinductor/width=1.5}
\def\BibTeX{{\rm B\kern-.05em{\sc i\kern-.025em b}\kern-.08em
    T\kern-.1667em\lower.7ex\hbox{E}\kern-.125emX}}
\usepackage[export]{adjustbox}
\usepackage{mathrsfs}
\def\BibTeX{{\rm B\kern-.05em{\sc i\kern-.025em b}\kern-.08em
    T\kern-.1667em\lower.7ex\hbox{E}\kern-.125emX}}
\usepackage{scalerel}
\usepackage{siunitx}
\usetikzlibrary{svg.path}
\definecolor{orcidlogocol}{HTML}{A6CE39}
\tikzset{
  orcidlogo/.pic={
    \fill[orcidlogocol] svg{M256,128c0,70.7-57.3,128-128,128C57.3,256,0,198.7,0,128C0,57.3,57.3,0,128,0C198.7,0,256,57.3,256,128z};
    \fill[white] svg{M86.3,186.2H70.9V79.1h15.4v48.4V186.2z}
                 svg{M108.9,79.1h41.6c39.6,0,57,28.3,57,53.6c0,27.5-21.5,53.6-56.8,53.6h-41.8V79.1z M124.3,172.4h24.5c34.9,0,42.9-26.5,42.9-39.7c0-21.5-13.7-39.7-43.7-39.7h-23.7V172.4z}
                 svg{M88.7,56.8c0,5.5-4.5,10.1-10.1,10.1c-5.6,0-10.1-4.6-10.1-10.1c0-5.6,4.5-10.1,10.1-10.1C84.2,46.7,88.7,51.3,88.7,56.8z};
  }
}
\newcommand\orcidicon[1]{\href{https://orcid.org/#1}{\mbox{\scalerel*{
\begin{tikzpicture}[yscale=-1,transform shape]
\pic{orcidlogo};
\end{tikzpicture}
}{|}}}}
\newcommand*{\field}[1]{\mathbb{#1}}

\begin{document}
\title{Interturn Short Circuit Fault Mitigation in PMSMs}

\author{
	\vskip 1em
	Lukas Zezula \orcidicon{0000-0002-3183-2438}, Matus Kozovsky \orcidicon{0000-0002-1547-1003}, Ludek Buchta \orcidicon{0000-0002-8954-3495} and Petr Blaha \orcidicon{0000-0001-5534-2065}
	
	\thanks{
	    This work was supported in part by the European Union through the project Robotics and Advanced Industrial Production under Grant CZ.02.01.01/00/22\_008/0004590, and in part by the Technological Agency of the Czech Republic through the project Center for Advanced Machines and Manufacturing Technology under Grant TN02000028. \emph{(Corresponding author: Lukas Zezula.)}
		
		The authors are with the CEITEC - Central European Institute of Technology, Brno University of Technology, 612 00 Brno, Czech Republic (e-mail: Lukas.Zezula@ceitec.vutbr).
        
		This work has been submitted to the IEEE for possible publication. Copyright may be transferred without notice, after which this version may no longer be accessible.
	}
}

\maketitle
	
\begin{abstract}
Interturn short circuits are among the most critical faults in permanent magnet synchronous motor drives, as they combine localized heating in the shorted stator phase with electrical asymmetry that distorts the current feedback used for torque-producing control. This article proposes a control-based mitigation method enabling the post-fault operation of standard three-phase motor drives without additional dedicated hardware. Using the diagnostic features inferred from standard control-loop signals, the method augments the field-oriented control structure with two mechanisms: resistive-loss-limited current-reference generation and reconstruction of the torque-producing current components in the feedback path. The reference generator is derived from a discrete-time post-fault model and minimizes resistive losses, whereas the feedback reconstruction provides fault-free torque-producing current components as controlled variables of the current loop. The experimental validation has demonstrated reductions of up to 23--36\% in the fault-induced increase in the input power and 18--27\% in the local segment-loss increase, while confirming real-time adaptation to progressive fault aggravation emulated by stepped changes in the short circuit resistance.
\end{abstract}

\begin{IEEEkeywords}
fault diagnosis, fault tolerant control, permanent magnet motors, short-circuit currents, fault currents, torque control.
\end{IEEEkeywords}

\markboth{}%
{}

\definecolor{limegreen}{rgb}{0.2, 0.8, 0.2}
\definecolor{forestgreen}{rgb}{0.13, 0.55, 0.13}
\definecolor{greenhtml}{rgb}{0.0, 0.5, 0.0}

\section{Introduction}
\IEEEPARstart{R}{ecently}, electric drives with permanent magnet synchronous motors (PMSMs) have been widely deployed in automated systems, ranging from autonomous vehicles to advanced industrial production lines. In such applications, conventional protection mechanisms that disable the drive after fault detection to prevent further damage (fail-safe approach) are often unacceptable, as the resulting loss of propulsion or actuation typically compromises the system-level availability \cite{FailDegraded_AV_Survey, SmartManuf_SelfHealing_Review} and, in many cases, also introduces safety risks \cite{SteerByWire_FailOperational_RESS}. Thus, modern electric drives are increasingly expected to remain operational after fault occurrence, providing either nominal performance in a fail-operational regime or reduced performance in a fail-degraded mode \cite{FailOp_Taxonomy}. Maintaining post-fault operation, unlike fail-safe protection, requires a comprehensive fault-handling chain that combines the timely detection of incipient faults, extraction of relevant fault features through fault diagnostics, and fault-tolerant control (FTC) aimed at preserving admissible operation while preventing further fault propagation \cite{PMSM_FD_FTC_Access, PMSM_FD_FTC_Integration_TTE, ITSC_Online_SOA_TEC, ITSC_AdaptiveFTC_TIM}.

Among PMSM faults, interturn short circuits (ISCs) are particularly critical because an initially minor insulation degradation between adjacent turns can rapidly escalate to a full phase-winding failure \cite{PMSM_FD_FTC_Access}. Once a shorted path is established, the back-electromotive force and inductive coupling generate a fault current \cite{ShortCircuitCurrent_Behavior_TPEL}, which produces localized resistive heating, accelerating the insulation degradation and promoting the fault propagation to the entire phase segment \cite{ITSC_Review_JESTPE, ITF_MTPL_TIE}. In addition to thermal degradation, the stator current imbalance caused by ISC appears as a torque ripple, degrading the drive performance \cite{ITSC_Review_JESTPE}. Consequently, ISC-oriented FTC must limit the fault-induced thermal loading by constraining the fault current or associated resistive losses \cite{ITSC_Online_SOA_TEC,ITF_MTPL_TIE} while suppressing the torque ripple to avoid drive-performance degradation \cite{PMSM_FD_FTC_Integration_TTE,DTC_SPMSM_ITSC_TorqueRipple_TPEL}.

Existing ISC-oriented FTC strategies can be classified into hardware-assisted and control-based approaches. In hardware-assisted FTC, motors with multiple three-phase subsystems are one of the central directions because their independently supplied winding sets can decouple the torque production in fault-free sets from the fault mitigation in the ISC-affected subsystem \cite{PMSM_FTC_Review_TTE_2025,TurnFault_CurrentInjection_Triple3P_TIE, DTPPMSM_ITSC_Compensation_IECON}. Similarly, converter modifications can enable active ISC mitigation by providing additional controllable current paths for terminal shorting, demagnetizing current injection, or flux-linkage reduction \cite{ITSC_MTPCC_TPEL, ITSC_Review_JESTPE}. A complementary route is to enhance the machine's structural tolerance to ISCs by improving the phase isolation, increasing the effective fault-loop inductance or impedance, reducing shorted-turn copper losses, or physically separating adjacent turns \cite{ASP_PMSM_ITSC_TEC,BifilarCoils_TTE}. Although hardware-assisted FTC can be very effective, it shifts the ISC mitigation toward a dedicated machine or converter design, thereby increasing the manufacturing costs, hardware complexity, and integration effort.

Control-based approaches, in contrast, aim to mitigate ISCs primarily at the algorithmic level by adapting the post-fault control objectives. For instance, current- and loss-oriented methods focus on limiting the thermal stress in the shorted turns: the safe operating area control derates the drive according to the estimated short circuit current \cite{ITSC_Online_SOA_TEC}, the maximum torque per loss control minimizes the total copper loss including the fault loop \cite{ITF_MTPL_TIE}, and the maximum torque per circulating current control constrains the fault current while preserving the torque capability \cite{ITSC_MTPCC_TPEL}. Moreover, compensation-oriented methods can address even the electromagnetic asymmetry caused by the fault: the diagnosis-integrated current injection uses zero-sequence-voltage features to reduce the torque ripple in field-oriented control \cite{PMSM_FD_FTC_Integration_TTE}, the torque injection and flux observer compensation extend this idea to direct torque control \cite{DTC_SPMSM_ITSC_TorqueRipple_TPEL}, the model predictive compensation introduces adaptive compensation currents into the post-fault cost function \cite{CommonPredictiveModel_ITF_TTE}, and the recent adaptive approaches use speed fluctuation or subspace current residuals to reduce the ripple \cite{ITSC_AdaptiveFTC_TIM,DTPPMSM_SubspaceResiduals_TPEL}.

Despite the recent progress, standard three-phase PMSM drives still lack a control-based ISC mitigation method to jointly limit fault-induced losses and suppress torque ripple using only the fault features from standard control-loop signals. This article develops a unified ISC mitigation framework that transforms real-time diagnostic estimates \cite{DTM_fault_diag, DTM_fault_curr_mon} acquired from standard control-loop signals into loss-constrained current references and torque-ripple compensation terms, while both the diagnostics and the mitigation are grounded in a universal post-ISC discrete-time model \cite{DTM, Thesis, DTM_PMSM_MPC_IECON}.

The rest of this article is organized as follows. Section~\ref{sec:fault_sign_analysis} analyzes the post-ISC behavior of the PMSM using the adopted discrete-time model and derives the resistive-loss formulation used for the mitigation. Section~\ref{sec:MTPA} develops the proposed ISC-aware \textit{dq} current reference generation strategy. Section~\ref{sec:integration} describes the integration of the proposed ISC mitigation within the field-oriented control structure. Section~\ref{sec:validation} validates the proposed method through laboratory experiments. Finally, Section~\ref{sec:conclusion} concludes the article.

\section{Analysis of Fault Signatures}
\label{sec:fault_sign_analysis}
This section discusses the influence of an interturn short circuit on the electrical quantities and the power delivered by a motor. The analysis proceeds from the discrete-time modeling of instantaneous post-ISC PMSM outputs (Section~\ref{subsec:post_fault_dynamics}) to a discussion of the steady-state behavior (Section~\ref{subsec:post_fault_behavior}).

\subsection{Post-ISC Dynamics in Discrete Time}
\label{subsec:post_fault_dynamics}

\begin{figure*}[b]\vspace{-0.5cm}\hrule
\begin{align}
\begin{bmatrix}
i_{d,h}(k+1) \\ i_{q,h}(k+1)
\end{bmatrix} &= e^{-\frac{R_s}{L_{eq}}T_s}\textbf{T}(k)\begin{bmatrix}
i_{d,h}(k) \\ i_{q,h}(k)
\end{bmatrix} + \frac{1-e^{-\frac{R_s}{L_{eq}}T_s}}{R_s}\textbf{T}(k)\begin{bmatrix}
u_d(k) \\ u_q(k)
\end{bmatrix} - \frac{\lambda_{pm}}{L_{eq}}\frac{\omega_e(k)}{\frac{R_s^2}{L_{eq}^2}+\omega_e^2(k)}\left(\textbf{I}-e^{-\frac{R_s}{L_{eq}}T_s}\textbf{T}(k)\right)\begin{bmatrix} \omega_e(k) \\ \frac{R_s}{L_{eq}}
\end{bmatrix}\label{eqn:DTMhealthy}\\
i_f(k+1) &= e^{-\frac{R_f}{L_f}T_s} i_f(k) + \frac{1-e^{-\frac{R_f}{L_f}T_s}}{R_f} \begin{bmatrix}\cos\bigl(\theta_e(k)+\phi\bigl) & -\sin\bigl(\theta_e(k)+\phi\bigl)\end{bmatrix}\begin{bmatrix} u_d(k) \\ u_q(k)\end{bmatrix}\label{eqn:DTMfault}\\
\begin{bmatrix}
i_d(k) \\ i_q(k)
\end{bmatrix} &= \begin{bmatrix}
i_{d,h}(k) \\ i_{q,h}(k)
\end{bmatrix} + \frac{2}{3}\frac{\sigma}{n_s}\begin{bmatrix}\cos\bigl(\theta_e(k)+\phi\bigl) \\ -\sin\bigl(\theta_e(k)+\phi\bigl)\end{bmatrix}i_f(k)
\label{eqn:DTMcoupling}
\end{align}
\end{figure*}
To analyze the impact of an ISC on the rotational reference frame currents $i_d$ and $i_q$, we adopted the simplified form of the discrete-time model presented in \cite{DTM}, which had been used for the fault diagnostics in \cite{DTM_fault_diag} and \cite{DTM_fault_curr_mon}. The model comprises three main equations: \eqref{eqn:DTMhealthy} captures the behavior of the fault-free \textit{dq} currents $i_{d,h}$ and $i_{q,h}$, \eqref{eqn:DTMfault} governs the fault current $i_f$, and \eqref{eqn:DTMcoupling} describes how $i_{d,h}$, $i_{q,h}$, and $i_f$ influence the total \textit{dq} currents. In \eqref{eqn:DTMhealthy}, the parameters $R_s$, $L_{eq}$, $\lambda_{pm}$, and $T_s$ stand for the comparable resistance of the stator phases, equivalent (or close) rotor reference frame inductance $L_{eq}\approx L_d\approx L_q$, permanent magnet flux linkage, and sampling period, respectively. A discrete-time step $k$-dependent variables $u_d(k)$ and $u_q(k)$ then represent the rotor reference frame voltages, while $\omega_e(k)$ and $\theta_e(k)$ are the electrical angular velocity and angle. Moreover, $\textbf{I}$ is the $2\times 2$ identity matrix and $\textbf{T}(k)$ denotes the rotation matrix defined as in
\begin{align}
\textbf{T}(k) = \begin{bmatrix}
\cos\bigl(T_s \omega_e(k)\bigl) & \sin\bigl(T_s \omega_e(k)\bigl) \\
-\sin\bigl(T_s \omega_e(k)\bigl) & \cos\bigl(T_s \omega_e(k)\bigl)
\end{bmatrix}.
\label{eqn:rot_mat}
\end{align}
The definition of fault current \eqref{eqn:DTMfault} then further includes $\phi$, $R_f$, and $L_f$, where $\phi = \{0,-2\pi/3,2\pi/3\}$, if ISC is in phase $\{a,b,c\}$, respectively, and $R_f$ and $L_f$ are the fault-related resistance and inductance modeled as in
\begin{align}
R_f &= n_p\left(1-\frac{\sigma}{n_s}\right)R_s+\frac{1}{3}\frac{\sigma}{n_s}R_s+\frac{n_s}{\sigma}R_{sc} \nonumber\\
L_f &= \frac{2}{3}\frac{\sigma}{n_s}n_p(n_s-1)L_{eq}+\frac{1}{3}\frac{\sigma}{n_s}\bigl(n_p(n_s-1)+1\bigl)L_0.
\label{eqn:fault_params}
\end{align}
In \eqref{eqn:fault_params}, $\sigma\in\langle0,1\rangle$ (fault severity) represents the portion of a concentrated stator winding segment shorted by $R_{sc}$, with $R_{sc}$ being the ISC resistance. In addition, the parameters $n_p$ and $n_s$ describe the stator winding configuration, where $n_p$ is the number of parallel branches and $n_s$ is the number of series-connected coils in each branch \cite{DTM}. Finally, $L_0$ denotes the zero-sequence inductance.

As shown in \eqref{eqn:DTMcoupling}, under fault-free motor operation ($\sigma=0$), the instantaneous \textit{dq} currents are determined solely by their healthy components $i_{d,h}$ and $i_{q,h}$. In contrast, when an ISC occurs, $i_d$ and $i_q$ are also affected by the fault current $i_f$; thus, $i_f$ impacts even the total electrical power $p_e = 3(i_du_d + i_qu_q)/2$ of a motor. The post-ISC electrical power was indirectly determined in \cite{DTM} and \cite{Thesis} during the energy analysis of a shorted motor, which was conducted to derive the electromagnetic torque $T_e$. When the fluctuation of stator inductances, higher harmonic components of the permanent magnet fluxes, and other nonlinearities (e.g., magnetic saturation and hysteresis) are ignored, the electrical power can be expressed at any given time $t = kT_s$ as in
\begin{align}
p_e(k) =&\, p_{res}(k) + \left.\diff{W_{mag}'(t)}{t}\right|_{t = kT_s} + p_{em}(k)\nonumber\\
p_{res}(k) =&\, \frac{3}{2}R_s\left(i_{d,h}^2(k)+i_{q,h}^2(k)\right) + \frac{\sigma}{n_s} R_f i_f^2(k) \nonumber\\
W_{mag}'(t) =&\, \frac{3}{4}L_{eq}\left(i_{d,h}^2(t)+i_{q,h}^2(t)\right) + \frac{3}{2} \lambda_{pm} i_{d,h}(t) \nonumber\\ 
&\,+ \frac{1}{2} \frac{\sigma}{n_s} L_f i_f^2(t) \nonumber\\
p_{em}(k) = &\, \frac{3}{2}\omega_e(k)\lambda_{pm}i_{q,h}(k).
\label{eqn:Electrical_power}
\end{align}
In \eqref{eqn:Electrical_power}, $p_{res}$ stands for the resistive power loss, $W_{mag}'$ is the magnetic co-energy, and $p_{em}$ represents the electromagnetic power. The electromagnetic torque is then derived from $p_{em}$ as follows:
\begin{align}
T_e(k) = \frac{P_P}{\omega_e(k)}p_{em}(k)= \frac{3}{2}P_P\lambda_{pm}i_{q,h}(k)
\label{eqn:torque}
\end{align}
where $P_P$ stands for the number of pole pairs. As shown in \eqref{eqn:torque}, when the fluctuation in the stator inductances and the higher-harmonic components of the permanent magnet fluxes are ignored, $T_e$ becomes directly proportional to the fault-free part of the quadrature axis current. Although $i_f$ has no direct influence on $T_e$, it affects the total \textit{dq} currents \eqref{eqn:DTMcoupling}, and the resulting distortion is propagated to the generated torque through the current control loop. In addition, as shown in \eqref{eqn:Electrical_power}, the presence of $i_f$ increases the resistive losses and magnetic co-energy. The rise of $p_{res}$ is then particularly problematic, as it often causes overheating that exceeds the motor's thermal limits. Finally, from \eqref{eqn:DTMcoupling} and \eqref{eqn:Electrical_power}, it can be concluded that minimizing the amplitude of $i_f$ reduces the adverse impact of an ISC on a motor operation.

As seen in \eqref{eqn:DTMfault}, the fault current embodies a harmonic signal with the instantaneous angular frequency $\omega_e(k)$. Consequently, $i_f$ causes a DC offset and a distortion at the second harmonics in the total \textit{dq} currents \eqref{eqn:DTMcoupling} and in $p_{res}$ and $W_{mag}'$ \eqref{eqn:Electrical_power}. The current $i_f$ can then be represented by its orthogonal components $i_{c,f}$ and $i_{s,f}$ as in 
\begin{align}
i_f(k) = i_{c,f}(k)\cos\bigl(\theta_e(k)+\phi\bigl) - i_{s,f}(k)\sin\bigl(\theta_e(k)+\phi\bigl)
\label{eqn:fault_curr_orthogonal1}
\end{align}
where the amplitude of $i_f$ reads $I_f=\sqrt{i_{c,f}^2+i_{s,f}^2}$. As mentioned in \cite{DTM}, the discrete-time post-ISC motor model is based on the assumption that the electrical angular velocity maintains an almost constant value $\omega_e(t)\approx \omega_e(k)$ over the sampling interval $kT_s\leq t<(k+1)T_s$, yielding a piece-wise linear angle $\theta_e(k+1)\approx \theta_e(k)+T_s\omega_e(k)$. This assumption then allows $i_f(k+1)$ to be written using the same orthogonal basis as in \eqref{eqn:fault_curr_orthogonal1}. We have
\begin{align}
i_f(k+1) = \left(\textbf{T}(k)\begin{bmatrix}
\cos\bigl(\theta_e(k)+\phi\bigl) \\ -\sin\bigl(\theta_e(k)+\phi\bigl)
\end{bmatrix}\right)^T\begin{bmatrix}i_{c,f}(k+1)\\i_{s,f}(k+1)\end{bmatrix}
\label{eqn:fault_curr_orthogonal2}
\end{align}
where $\textbf{T}(k)$ corresponds to \eqref{eqn:rot_mat}. Finally, substituting \eqref{eqn:fault_curr_orthogonal1} and \eqref{eqn:fault_curr_orthogonal2} into \eqref{eqn:DTMfault}, followed by grouping the coefficients of the sine and cosine functions, yields
\begin{align}
&\begin{bmatrix}i_{c,f}(k+1)\\i_{s,f}(k+1)\end{bmatrix} = \nonumber\\ &e^{-\frac{R_f}{L_f}T_s} \textbf{T}(k)\begin{bmatrix}i_{c,f}(k)\\i_{s,f}(k)\end{bmatrix} + \frac{1-e^{-\frac{R_f}{L_f}T_s}}{R_f}\textbf{T}(k)\begin{bmatrix}
u_d(k) \\ u_q(k)
\end{bmatrix}. 
\label{eqn:DTMfault_orthogonal}
\end{align}
\begin{figure*}[b]\vspace{-0.5cm}\hrule
\begin{align}
\sqrt{u_d^2+u_q^2} = \frac{R_s\sqrt{\left(1-e^{-\frac{R_s}{L_{eq}}T_s}\right)^2+2e^{-\frac{R_s}{L_{eq}}T_s}\bigl(1-\cos(T_s\omega_e)\bigl)}}{1-e^{-\frac{R_s}{L_{eq}}T_s}}\sqrt{\left(i_{d,h}+\frac{\lambda_{pm}}{L_{eq}}\frac{\omega_e^2}{\frac{R_s^2}{L_{eq}^2}+\omega_e^2}\right)^2+\left(i_{q,h}+\frac{\lambda_{pm}}{L_{eq}}\frac{\omega_e\frac{R_s}{L_{eq}}}{\frac{R_s^2}{L_{eq}^2}+\omega_e^2}\right)^2}
\label{eqn:volt_mag}
\end{align}
\end{figure*}
\par\noindent Model \eqref{eqn:DTMfault_orthogonal} then enables predicting $I_f$ in each discrete step $k$. Importantly, as shown in \eqref{eqn:DTMfault_orthogonal}, the difference equations governing $i_{c,f}$ and $i_{s,f}$ closely resemble the behavior of the fault-free \textit{dq} currents \eqref{eqn:DTMhealthy}; thus, in response to the constant \textit{dq} voltages, $i_{c,f}$ and $i_{s,f}$ settle at steady values, resulting in a fixed amplitude of the fault current. A more detailed discussion of the system’s steady-state behavior is presented next.

\subsection{Steady-State Post-ISC Behavior}
\label{subsec:post_fault_behavior}
As discussed in Section~\ref{subsec:post_fault_dynamics}, the ISC-driven distortion in the total \textit{dq} currents \eqref{eqn:DTMcoupling} is propagated through the feedback loop and affects the voltage control actions. Yet the electromagnetic torque is determined solely by $i_{q,h}$. It is, therefore, advantageous to subtract the influence of $i_f$ from $i_d$ and $i_q$ within the field-oriented control (FOC) structure. Accordingly, in the steady-state analysis, we assume that the distortion terms are not propagated through the control loop. Consequently, the \textit{dq} voltages settle to constants $u_d(k+n)=u_d(k)=u_d$ and $u_q(k+n)=u_q(k)=u_q$ for $n\in\field{N}$ in the velocity steady state, characterized by $\omega_e(k+n)=\omega_e(k)=\omega_e$, yielding stabilized $i_{d,h}$ and $i_{q,h}$. The voltage vector magnitude $\sqrt{u_d^2+u_q^2}$ necessary to attain the values $i_{d,h}$ and $i_{q,h}$ can then be derived from \eqref{eqn:DTMhealthy}, as expressed in \eqref{eqn:volt_mag}, shown at the bottom of the page. By employing the approximations $1-\cos(T_s\omega_e)\approx T_s^2\omega_e^2/2$ and $e^{\frac{1}{2}\frac{R_s}{L_{eq}}T_s}-e^{-\frac{1}{2}\frac{R_s}{L_{eq}}T_s}\approx T_sR_s/L_{eq}$, the voltage vector magnitude reduces to the well-known expression
\begin{align}
\sqrt{u_d^2+u_q^2} &\approx \sqrt{c_2(i_{d,h},i_{q,h})\omega_e^2+c_1(i_{q,h})\omega_e + c_0(i_{d,h},i_{q,h})}
\label{eqn:volt_mag2}
\end{align}
where $c_2(i_{d,h},i_{q,h}) = (L_{eq}i_{d,h}+\lambda_{pm})^2+L_{eq}^2i_{q,h}^2$, $c_1(i_{q,h})=2R_s\lambda_{pm}i_{q,h}$, and $c_0(i_{d,h},i_{q,h}) = R_s^2(i_{d,h}^2+i_{q,h}^2)$. Expectably, an increase in the torque load or the electrical angular velocity leads to the rise of $\sqrt{u_d^2+u_q^2}$. However, such a rise also elevates the orthogonal components of the fault current, which converge to the following values:
\begin{align}
\begin{bmatrix}i_{c,f}\\i_{s,f}\end{bmatrix} = \left(\textbf{I}-e^{-\frac{R_f}{L_f}T_s} \textbf{T}\right)^{-1} \frac{1-e^{-\frac{R_f}{L_f}T_s}}{R_f}\textbf{T}\begin{bmatrix}
u_d \\ u_q
\end{bmatrix}
\label{eqn:DTMfault_orthogonal_steady}
\end{align}
where $\textbf{T}$ is the rotation matrix \eqref{eqn:rot_mat} with the stable velocity $\omega_e(k+n)=\omega_e(k)=\omega_e$. The settled fault current amplitude $I_f=\sqrt{i_{c,f}^2+i_{s,f}^2}$ then reads
\begin{align}
I_f &= \frac{\left(1-e^{-\frac{R_f}{L_f}T_s}\right)\sqrt{u_d^2+u_q^2}}{R_f\sqrt{\left(1-e^{-\frac{R_f}{L_f}T_s}\right)^2 + 2e^{-\frac{R_f}{L_f}T_s}\bigl(1-\cos(T_s\omega_e)\bigl)}}\nonumber\\
I_f &\approx \frac{K_f}{R_f}\frac{\sqrt{u_d^2+u_q^2}}{\sqrt{K_f^2+\omega_e^2}} \qquad K_f = \frac{e^{\frac{1}{2}\frac{R_f}{L_f}T_s}-e^{-\frac{1}{2}\frac{R_f}{L_f}T_s}}{T_s}.
\label{eqn:DTMfault_curr_amplitude_steady}
\end{align}
A first-order Taylor expansion of $K_f$ is not adopted, because low-severity ISCs ($\sigma\to0$) result in $R_f\to\infty$ and $L_f\to0$, making the approximation insufficiently accurate. As shown in \eqref{eqn:DTMfault_curr_amplitude_steady}, the fault current amplitude is directly proportional to the voltage vector magnitude and can be, therefore, parameterized as $I_f(i_{d,h}, i_{q,h}, \omega_e)$. The observed proportionality suggests a mitigation strategy that minimizes $I_f$ by steering the operating point toward reduced voltage consumption, i.e., the maximum torque per volt (MTPV) trajectory. However, under the hard current limit $\sqrt{i_{d,h}^2+i_{q,h}^2}\leq I_{\max}$, the resulting operating points can become excessively restrictive in torque production. Moreover, reducing $I_f$ does not automatically guarantee thermal stress reduction, because the ISC-related loss in \eqref{eqn:Electrical_power} is lowered only at the cost of increased fault-free copper losses through $i_{d,h}$. Thus, the thermal load is redistributed instead of being mitigated. Consequently, the fault mitigation objective should be aligned with the dominant risk, i.e., overheating, and should minimize the heat losses rather than $I_f$.

Within the adopted model, described in Section~\ref{subsec:post_fault_dynamics}, the thermal loading is quantified through the term $p_{res}$, which represents the resistive losses under ISC conditions. In a velocity steady state, the fault current is sinusoidal, and its thermal effect is most naturally expressed via the root-mean-square value $I_f/\sqrt{2}$, yielding
\begin{align}
p_{res} = \frac{3}{2}R_s\left(i_{d,h}^2+i_{q,h}^2\right) + \frac{1}{2}\frac{\sigma}{n_s} R_f I_f^2(\omega_e, i_{d,h}, i_{q,h}).
\label{eqn:pres_steady}
\end{align}
In the nominal motor operation, the reference currents track the maximum torque per ampere (MTPA) trajectory (i.e., $i_{d,h}=0$) until the voltage limit $U_{\max}$ is reached, beyond which flux weakening (FW) is initiated. Under these conditions, the worst-case resistive loss $p_{res}^{\max}$ occurs when the maximum torque is generated (i.e, $i_{q,h}=I_{\max}$) at $\omega_e$ corresponding to $\omega_e^\star =\arg\max_{\omega_e} p_{res}(i_{d,h}=0, i_{q,h}=I_{\max}, \omega_e)$, where
\begin{align}
\omega_e^\star &= \omega_{e,0}^\star + \sqrt{{\omega_{e,0}^\star}^2 + K_f^2} \nonumber\\
\omega_{e,0}^\star &= \frac{\lambda_{pm}^2+L_{eq}^2I_{\max}^2}{2R_s\lambda_{pm}I_{\max}}K_f^2 - \frac{R_s}{2}\frac{I_{\max}}{\lambda_{pm}}.
\label{eqn:presmax_velocity}
\end{align}
Importantly, at low fault severities, $K_f$ can increase dramatically, shifting $\omega_e^\star$ towards very high frequencies. However, such $\omega_e^\star$ typically falls outside the feasible operating region when the voltage and speed limits are enforced. In such a case, as evident from \eqref{eqn:pres_steady} and \eqref{eqn:DTMfault_curr_amplitude_steady}, $p_{res}^{\max}$ is achieved at the voltage limit boundary, i.e., $\sqrt{u_{d}^2+u_{q}^2}= U_{\max}$, where $\omega_e=\omega_b$. We have
\begin{align}
\omega_b = \frac{\sqrt{(\lambda_{pm}^2+L_{eq}^2I_{\max}^2)U_{\max}^2-L_{eq}^2R_s^2I_{\max}^4}-R_s\lambda_{pm}I_{\max}}{\lambda_{pm}^2+L_{eq}^2I_{\max}^2}
\label{eqn:base_speed}
\end{align}
where $\omega_b$ denotes the base speed of a motor. Finally, the maximum resistive loss reads
\begin{align}
p_{res}^{\max} &= \frac{3}{2}R_s I_{\max}^2 + \frac{1}{2}\frac{\sigma}{n_s}R_f\left(I_f^{\max}\right)^2\nonumber\\
I_f^{\max} &= 
\begin{cases}\frac{K_f}{R_f} \sqrt{\frac{L_{eq}^2I_{\max}^2{\omega_e^\star}^2 + \left(\lambda_{pm}\omega_e^\star+R_sI_{\max}\right)^2}{K_f^2+{\omega_e^\star}^2}} & \omega_e^\star<\omega_b \\
\frac{K_f}{R_f}\frac{U_{\max}}{\sqrt{K_f^2+\omega_b^2}} & \omega_e^\star\geq\omega_b
\end{cases} 
\label{eqn:presmax_value}
\end{align}
As shown in \eqref{eqn:presmax_value}, for $\sigma\to0$ the worst-case resistive loss $p_{res}^{\max}$ reduces to $3R_sI_{\max}^2/2$, a limit set from the thermal considerations of a fault-free motor. However, if $\sigma>0$, $p_{res}^{\max}$ is elevated by an additional fault current-dependent term, and the nominal thermal envelope can be violated unless the current references are adapted. Accordingly, we pose fault mitigation as a thermal-loading problem: the reference currents $i_{d,h}$ and $i_{q,h}$ must ensure $p_{res}\leq3R_sI_{\max}^2/2$ while meeting the torque demands whenever feasible. Minimizing $p_{res}$ at fixed $T_e$ and $\omega_e$ then formally yields an ISC-aware MTPA curve. 

\section{ISC-aware MTPA reference}
\label{sec:MTPA}
If the resistive loss limitation $p_{res}\leq 3R_sI_{\max}^2/2$ is enforced, an additional circular constraint arises. We have
\begin{align}
r^2(\omega_e) &\geq \bigl(i_{d,h} - i_{d,h,0}(\omega_e)\bigl)^2 + \bigl(i_{q,h} - i_{q,h,0}(\omega_e)\bigl)^2\nonumber\\
r(\omega_e) &= \sqrt{\frac{I_{\max}^2}{1+G_f(\omega_e)}-\frac{\lambda_{pm}^2}{L_{eq}^2}\frac{G_f(\omega_e)}{\bigl(1+G_f(\omega_e)\bigl)^2}\frac{\omega_e^2}{\frac{R_{s}^2}{L_{eq}^2} + \omega_e^2}}\nonumber\\
i_{d,h,0}(\omega_e) &= -\frac{\lambda_{pm}}{L_{eq}}\frac{G_f(\omega_e)}{1+G_f(\omega_e)}\frac{\omega_e^2}{\frac{R_{s}^2}{L_{eq}^2} + \omega_e^2}\nonumber\\
i_{q,h,0}(\omega_e) &= -\frac{\lambda_{pm}}{L_{eq}}\frac{G_f(\omega_e)}{1+G_f(\omega_e)}\frac{\omega_e\frac{R_{s}}{L_{eq}}}{\frac{R_{s}^2}{L_{eq}^2} + \omega_e^2}\nonumber\\
G_f(\omega_e) &= \frac{1}{3}\frac{\sigma}{n_s}\frac{R_s}{R_f}\frac{1+\frac{L_{eq}^2}{R_s^2}\omega_e^2}{1+\frac{1}{K_f^2}\omega_e^2}.
\label{eqn:res_loss_limit}
\end{align}
Importantly, if $\sigma\to0$ or $R_{sc}\to\infty$, then $G_f\to0$ and constraint \eqref{eqn:res_loss_limit} reduce to the current limit $I_{\max} \geq \sqrt{i_{d,h}^2+i_{q,h}^2}$. Conversely, the most restrictive circle does not necessarily result from the extreme case $\sigma\to1$ and $R_{sc}\to0$. Instead, it is attained by maximizing $G_f$ over the admissible set $(\sigma, R_{sc})$ at a fixed $\omega_e$. The maximization can then be re-parameterized in terms of the ratio $R_f/L_f$. For a fixed $\omega_e$, $G_f$ is expressible as $G_f(R_f/L_f)$, while $R_f/L_f$ is induced by the combination of $\sigma$ and $R_{sc}$, where infinitely many distinct pairs $(\sigma, R_{sc})$ can result in the same ratio $R_f/L_f$. Moreover, the maximizer is not anticipated in the low-severity limit $(\sigma\to1, R_{sc}\to0)$, where $K_f\to\infty$ and $G_f\to0$. Thus, we employ the approximation $K_f\approx R_f/L_f$, which yields $\arg\max_{R_f/L_f}G_f = \omega_e$. Having the maximizing $R_f/L_f$ derived, the dependency of $G_f$ on $\omega_e$ can be characterized using the following asymptotes:
\begin{align}
G_f(\omega_e) \approx \begin{cases} \frac{1}{3} \frac{\sigma}{n_s}\frac{L_{eq}}{L_{f}}\frac{R_f}{L_f}\frac{L_{eq}}{R_s} & \frac{R_f}{L_f}<<\omega_e \\
\frac{1}{3}\frac{\sigma}{n_s}\frac{L_{eq}}{L_{f}}\left(\frac{1}{\frac{L_{eq}}{R_s}\omega_e} + \frac{L_{eq}}{R_s}\omega_e\right)\frac{1}{2} & \frac{R_f}{L_f}\to\omega_e \\
\frac{1}{3} \frac{\sigma}{n_s}\frac{L_{eq}}{L_{f}}\left(\frac{1}{\frac{L_{eq}}{R_s}\omega_e} + \frac{L_{eq}}{R_s}\omega_e\right)\frac{L_f}{R_f}\omega_e & \frac{R_f}{L_f}>>\omega_e.\end{cases} 
\label{eqn:Gf_approx}
\end{align}
As shown in \eqref{eqn:Gf_approx}, for $R_f/L_f<<\omega_e$, $G_f$ approaches the constant value. By contrast, when $R_f/L_f\to\omega_e$ or $R_f/L_f>>\omega_e$, $G_f(\omega_e)$, in most practical cases, grows as the angular velocity increases. A decrease of $G_f(\omega_e)$ with growing $\omega_e$ is only obtained in the special case $R_f/L_f\approx\omega_e<R_s/L_{eq}$.

\begin{figure*}[t]
\centering
\includegraphics[]{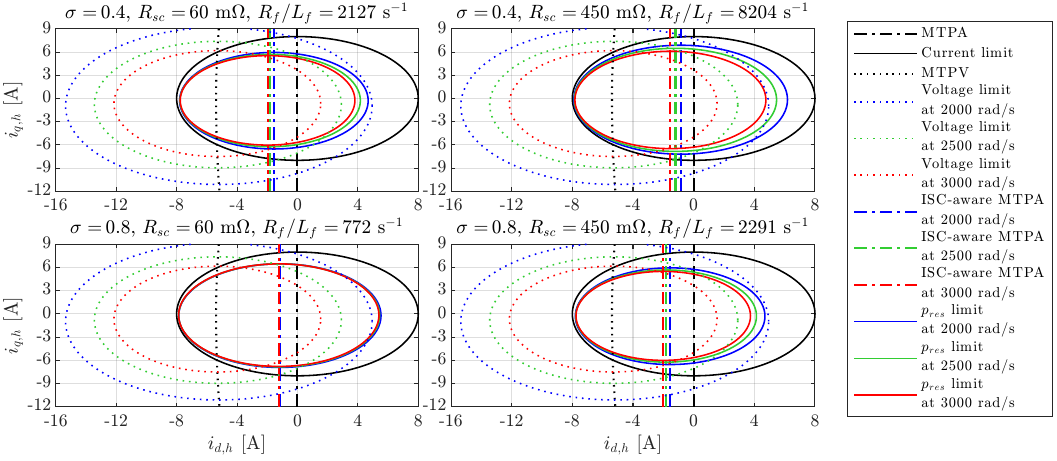}
\vspace{-0.4cm}
\caption{The post-ISC constraint curves and ISC-aware MTPA for diverse $\sigma$, $R_{sc}$, and $\omega_e$.}\vspace{-0.3cm}
\label{fig:Constr_curves}
\end{figure*}

\begin{figure*}[t]
\centering
\includegraphics[width=\textwidth]{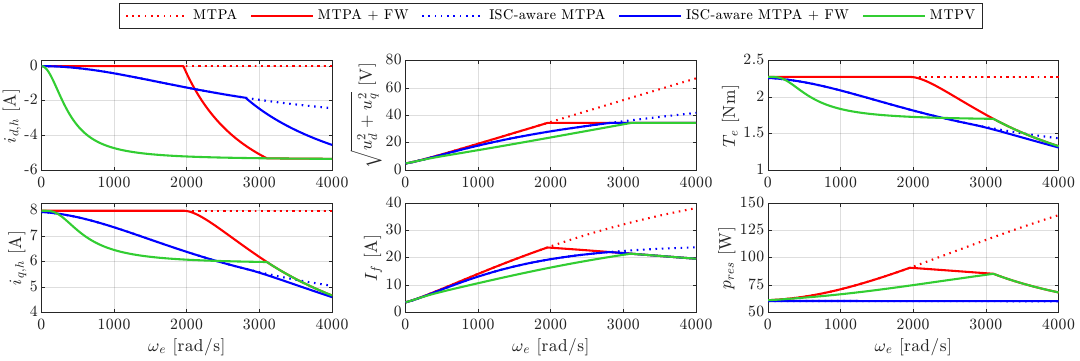}
\vspace{-0.8cm}
\caption{A comparison of the reference trajectories under $\sigma=6/25$, $R_{sc}=60$ \si{\milli\ohm} ISC across $\omega_e$ at the maximum achievable $T_e$.}\vspace{-0.3cm}
\label{fig:If_behav}
\end{figure*}

Having determined the admissible current set, we translate the constraint \eqref{eqn:res_loss_limit} into a reference-generation rule. Specifically, $i_{d,h}$ is chosen to minimize $p_{res}$ at the required torque (ISC-aware MTPA), as in
\begin{align}
i_{d,h}^{MTPA} = \arg\min_{i_{d,h}} p_{res} (i_{d,h}, i_{q,h}, \omega_e) = i_{d,h,0}(\omega_e)
\label{eqn:MTPA}
\end{align}
where $i_{d,h,0}(\omega_e)$ represents the \textit{d}-axis center of the limit circle \eqref{eqn:res_loss_limit}. Although $i_{d,h}^{MTPA}$ is already shifted toward negative values compared to the fault-free case, the corresponding value $i_{d,h,0}(\omega_e)$ does not necessarily guarantee compliance with the voltage constraint in the high-speed region. Consequently, the effective \textit{d}-axis current reference $i_{d,h}^\star = i_{d,h}^{MTPA} + i_{d,h}^{FW}$ is augmented by an additional component, $i_{d,h}^{FW}$, generated by the field-weakening controller. Given $i_{d,h}^\star$, the quadrature current reference $i_{q,h}^\star$ limit follows directly from \eqref{eqn:res_loss_limit}. We have
\begin{align}
i_{q,h}^\star &\leq i_{q,h,0}(\omega_e) + \sqrt{r^2(\omega_e)-\bigl(i_{d,h}^\star - i_{d,h,0}(\omega_e)\bigl)^2}\nonumber\\
i_{q,h}^\star &\geq i_{q,h,0}(\omega_e) - \sqrt{r^2(\omega_e)-\bigl(i_{d,h}^\star - i_{d,h,0}(\omega_e)\bigl)^2}.
\label{eqn:iq_limit}
\end{align}
For completeness, in the high-speed region, $i_{d,h}^\star$ is saturated on the MTPV boundary, whereas $i_{q,h}^\star$ is limited by the remaining voltage margin \eqref{eqn:volt_mag2}.

Figure~\ref{fig:Constr_curves} illustrates how the current, voltage, and resistive-loss constraints jointly shape the post-ISC feasible current region. The constraint curves were plotted using the estimated parameters of the experimental motor, which are detailed in Section~\ref{sec:validation}. In agreement with the above analysis, the resistive loss circles are most restrictive for $R_f/L_f\to\omega_e$ and tighten with increasing $\omega_e$. In addition, for $R_f/L_f<<\omega_e$, the circles become nearly independent of $\omega_e$, and for $R_f/L_f>>\omega_e$, the resistive-loss constraint becomes less restrictive. Furthermore, different $(\sigma,R_{sc})$ lead to a nearly equivalent behavior, as they induce comparable $R_f/L_f$ ratios and thus similar thermal effects. Figure~\ref{fig:If_behav} then shows the reference trajectories for the considered ISC case ($\sigma=6/25$, $R_{sc}=60$~\unit{\milli\ohm}) at the maximum achievable $T_e$. Unlike the conventional MTPA and MTPV trajectories, the proposed ISC-aware MTPA also enforces the resistive-loss limit, thereby maintaining $p_{res}$ at the prescribed maximum. Interestingly, compared with the MTPV control at a lower speed, the proposed trajectory does not suppress the fault current as aggressively, but it preserves a higher torque capability. The reason is that the ISC-aware MTPA does not minimize $I_f$ alone; instead, it balances the thermal effect of the fault current with the natural stator copper losses. As a result, elevated \textit{d}-axis currents are avoided because their contribution to the copper-loss term in \eqref{eqn:pres_steady} offsets the benefit of further fault-current reduction. Conversely, at elevated speeds, satisfying the prescribed maximum $p_{res}$ requires a reduction in the maximum allowable torque compared with MTPA and MTPV operation.

\section{Integration of the fault mitigation strategy}
\label{sec:integration}
\begin{figure*}
\input{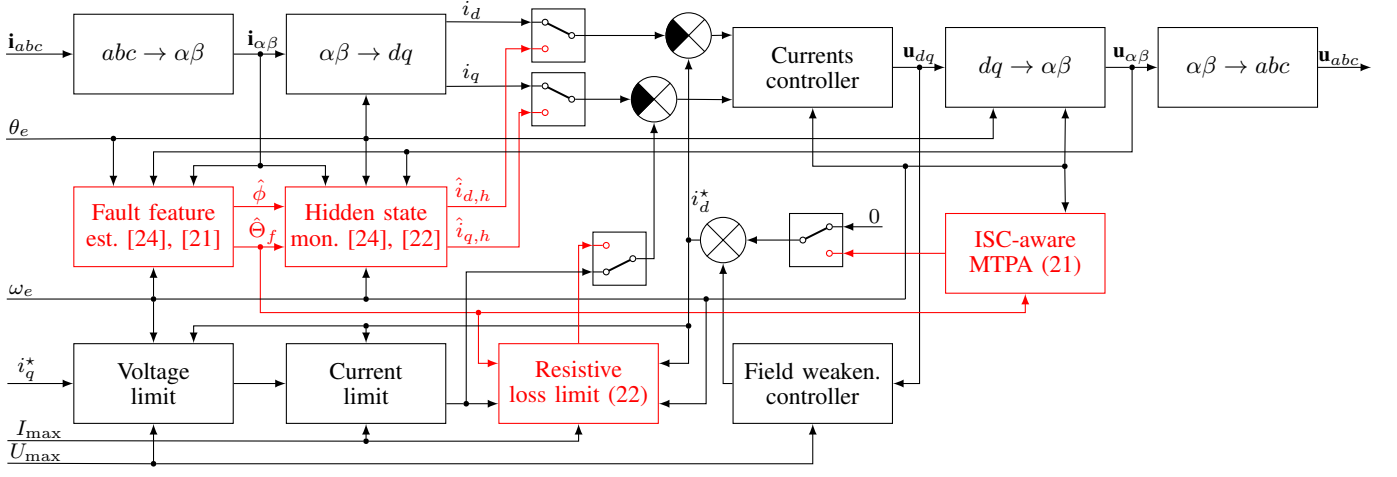}
\vspace{-0.5cm}
\caption{The integration of the proposed fault mitigation into the field-oriented control structure.}\vspace{-0.3cm}
\label{fig:Fault_mit_integration}
\end{figure*}
The manner in which the designed ISC mitigation integrates within the FOC structure is visualized in Fig.~\ref{fig:Fault_mit_integration}. Here, the reference \textit{dq} currents $i_d^\star$ and $i_q^\star$ are modified according to \eqref{eqn:MTPA} and \eqref{eqn:iq_limit} to track the ISC-aware MTPA and satisfy the imposed resistive-loss constraint. Importantly, evaluating \eqref{eqn:MTPA} and \eqref{eqn:iq_limit} requires prior computation of the $G_f$-dependent terms in \eqref{eqn:res_loss_limit}. As $G_f$ is parameterized by $\sigma$, $R_f$, and $L_f$, its value has to be inferred using post-detection fault diagnostics. To evaluate $G_f$, we employ the outputs of the fault diagnostics originally introduced in \cite{DTM_fault_diag} and subsequently refined and detailed in \cite{Thesis}. Using only signals available within the control structure, the diagnostic framework recursively adapts the estimates of $R_s$ and $L_{eq}$ to the current operating point, performs uncertainty-aware fault detection, and enables post-detection ISC localization and fault-severity estimation. Specifically, during the fault severity identification phase, the diagnostic algorithm recursively updates the estimates $\hat{\Theta}_f = \begin{bmatrix} \hat{\theta}_{f,1}(k) & \hat{\theta}_{f,2}(k)\end{bmatrix}$ of the following ISC-related parameters:
\begin{align}
\Theta_f = \begin{bmatrix} e^{-\frac{R_f}{L_f}T_s} & \frac{2}{3}\frac{\sigma}{n_s}\frac{1-e^{-\frac{R_f}{L_f}T_s}}{R_f}\end{bmatrix}^T.
\label{eqn:ISC_params}
\end{align}
Using the estimated values $\hat{\Theta}_f$, the identified $G_f$ (i.e., $\hat{G}_f(k)$) can be evaluated in each discrete step $k$, as in
\begin{align}
\hat{G}_f(k) = \frac{1}{2}  \frac{\bigl(1-\hat{\theta}_{f,1}(k)\bigl)\hat{\theta}_{f,2}(k)R_s\left(1+\frac{L_{eq}^2}{R_s^2}\omega_e^2(k)\right)}{\bigl(1-\hat{\theta}_{f,1}(k)\bigl)^2 + \hat{\theta}_{f,1}(k) T_s^2\omega_e^2(k)}.
\label{eqn:GF_ISC_params}
\end{align}
Finally, $\hat{G}_f(k)$ is employed in \eqref{eqn:res_loss_limit}, \eqref{eqn:MTPA}, \eqref{eqn:iq_limit} to construct the ISC-aware \textit{dq} current references. 

As also shown in Fig.~\ref{fig:Fault_mit_integration}, the current controller operates on the fault-free \textit{dq} current estimates instead of $i_d$ and $i_q$. Variables $\hat{i}_{d,h}$ and $\hat{i}_{q,h}$ are then recursively updated using the Kalman filter-based algorithm \cite{DTM_fault_curr_mon, Thesis}, which is driven by the control structure signals and parameterized by the fault indicator $\hat{\Theta}_f$ and shorted-phase location $\hat{\phi}$ estimates obtained from the diagnostics \cite{DTM_fault_diag}. Specifically, the monitoring algorithm provides estimates $\hat{\textbf{x}}(k) = \begin{bmatrix} \hat{i}_{\alpha,h}(k) &  \hat{i}_{\beta,h}(k) & \hat{\bar{i}}_f(k)\end{bmatrix}$ together with the updated covariance matrix $\textbf{P}(k)$. The fault-free $\alpha\beta$ currents are then transformed to the \textit{dq} frame and used in the current controller. The variable $\bar{i}_f=(2/3)(\sigma/n_s)i_f$ denotes the fault-current projection and is estimated instead of $i_f$, as the fault current amplitude cannot be determined exactly using only the signals available in the control structure \cite{DTM_fault_diag, DTM_fault_curr_mon, Thesis}. This limitation arises because $\sigma$ and $R_{sc}$ are not separately identifiable from the available measurements, allowing only the aggregate parameters $R_f/L_f$ and $R_f/\sigma$ to be determined. Importantly, the fault mitigation is not activated immediately upon the fault detection. As the designed FTC strategy relies on the identified variables, the mitigation is postponed until the corresponding estimates have sufficiently stabilized and reached the required accuracy. The switching interface for the switches in Fig.~\ref{fig:Fault_mit_integration} is then defined based on the trace of the covariance matrix \textbf{P}(k). Specifically, the mitigation is enabled only when the variances of the individual estimates in $\hat{\textbf{x}}(k)$ fall below a threshold $\xi$, i.e., when $\text{tr}(\textbf{P}(k))\leq\xi$.

\section{Validation}
\label{sec:validation}

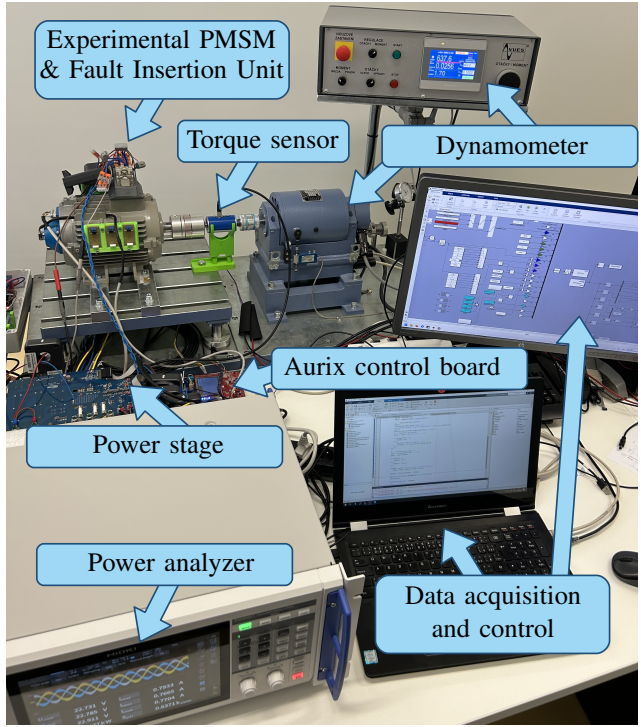
\begin{figure}[!t]
    \centering
    \def\svgwidth{\dimexpr\columnwidth-0.5cm\relax}
\begingroup%
  \makeatletter%
  \providecommand\color[2][]{%
    \errmessage{(Inkscape) Color is used for the text in Inkscape, but the package 'color.sty' is not loaded}%
    \renewcommand\color[2][]{}%
  }%
  \providecommand\transparent[1]{%
    \errmessage{(Inkscape) Transparency is used (non-zero) for the text in Inkscape, but the package 'transparent.sty' is not loaded}%
    \renewcommand\transparent[1]{}%
  }%
  \providecommand\rotatebox[2]{#2}%
  \newcommand*\fsize{\dimexpr\f@size pt\relax}%
  \newcommand*\lineheight[1]{\fontsize{\fsize}{#1\fsize}\selectfont}%
  \ifx\svgwidth\undefined%
    \setlength{\unitlength}{539.3074865bp}%
    \ifx\svgscale\undefined%
      \relax%
    \else%
      \setlength{\unitlength}{\unitlength * \real{\svgscale}}%
    \fi%
  \else%
    \setlength{\unitlength}{\svgwidth}%
  \fi%
  \global\let\svgwidth\undefined%
  \global\let\svgscale\undefined%
  \makeatother%
  \begin{picture}(1,1.14257636)%
    \lineheight{1}%
    \setlength\tabcolsep{0pt}%
    \put(0,0){\includegraphics[width=\unitlength,page=1]{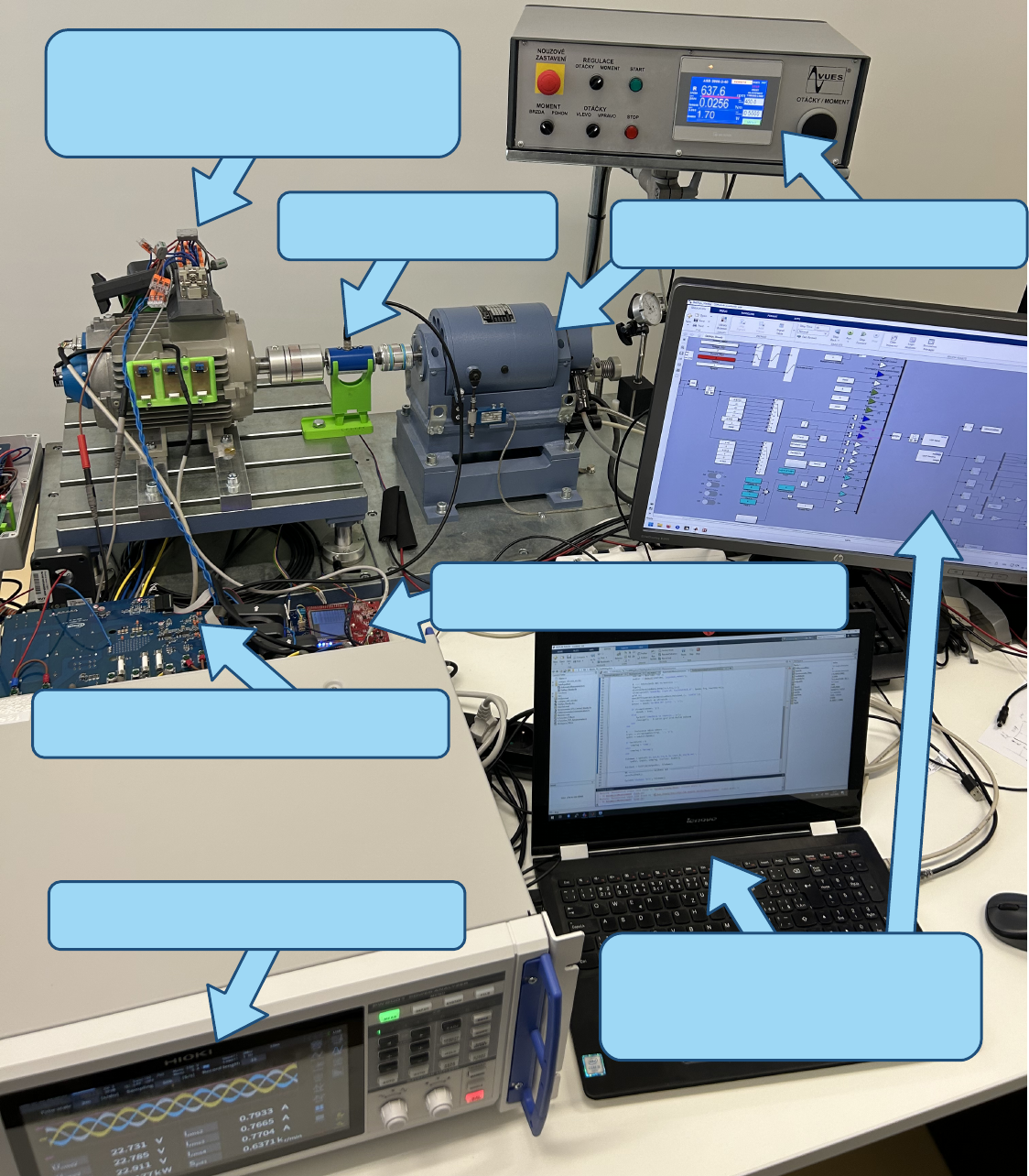}}%
    \put(0.77,0.19){\color[rgb]{0,0,0}\makebox(0,0)[cc]{\lineheight{1.25}\smash{\begin{tabular}[t]{c}Data acquisition\\and control\end{tabular}}}}%
  
    \put(0.405,0.913){\color[rgb]{0,0,0}\makebox(0,0)[cc]{\lineheight{1.25}\smash{\begin{tabular}[t]{c}Torque sensor\end{tabular}}}}%
    \put(0.80,0.903){\color[rgb]{0,0,0}\makebox(0,0)[cc]{\lineheight{1.25}\smash{\begin{tabular}[t]{c}Dynamometer\end{tabular}}}}%
    \put(0.61,0.553){\color[rgb]{0,0,0}\makebox(0,0)[cc]{\lineheight{1.25}\smash{\begin{tabular}[t]{c}Aurix control board\end{tabular}}}}%
    \put(0.24,0.43){\color[rgb]{0,0,0}\makebox(0,0)[cc]{\lineheight{1.25}\smash{\begin{tabular}[t]{c}Power stage\end{tabular}}}}%
    \put(0.26,0.242){\color[rgb]{0,0,0}\makebox(0,0)[cc]{\lineheight{1.25}\smash{\begin{tabular}[t]{c}Power analyzer\end{tabular}}}}%
    \put(0.245,1.07){\color[rgb]{0,0,0}\makebox(0,0)[cc]{\lineheight{1.125}\smash{\begin{tabular}[t]{c}Experimental PMSM\\\& Fault Insertion Unit\end{tabular}}}}%
  \end{picture}%
\endgroup%
    \vspace{-0.2cm}
    \caption{The testbench with the experimental PMSM.}
    \label{Fig:TestBench}
    \vspace{-0.3cm}
\end{figure}

The proposed ISC mitigation strategy was experimentally validated on the PMSM testbench shown in Fig.~\ref{Fig:TestBench}. The motor has accessible winding taps, allowing a selected portion of the stator winding segment to be externally shorted without damaging the machine. The short circuits are then emulated by a fault insertion unit (FIU), which connects the tapped winding segment through one of four digitally selectable resistance levels ($R_{sc,1}$--$R_{sc,4}$). Further details on the FIU are provided in \cite{DTM}. By switching to progressively lower $R_{sc}$ values, the FIU can emulate an increasingly severe ISC under controlled, repeatable conditions. 

\begin{table}[t]
	\caption{The experimental setup parameters}\vspace{-0.4cm}
	\label{tab:params}
    \setlength\tabcolsep{3pt}
    \def\arraystretch{1.2}
    \begin{center}
	\begin{tabular}{|p{1.6cm}|p{0.85cm}||p{1.6cm}|p{0.85cm}||p{1.45cm}|p{0.85cm}|}
		\hline
		Parameter & Value & Parameter & Value & Parameter & Value \\
		\hline
        $R_s$ [\unit{\milli\ohm}] & 630 & $n_p$ [-] & 1 & $\sigma$ [-] & $6/25$\\
        \hline
        $\lambda_{pm}$ [\unit{\milli\weber}] & 9.05 & $n_s$ [-] & 3 & $R_{sc,1}$ [\unit{\milli\ohm}] & 456 \\
        \hline
        $L_{eq}$ [\unit{\milli\henry}] & 1.68 & $I_{\max}$ [\unit{\ampere}] & 8 & $R_{sc,2}$ [\unit{\milli\ohm}] & 61.4 \\
        \hline
        $L_0$ [\unit{\milli\henry}] & 1.37 & $U_{\max}$ [\unit{\volt}] & $60/\sqrt{3}$ & $R_{sc,3}$ [\unit{\milli\ohm}] & 20.0\\
        \hline 
         $P_P$ [-] & 21 & $\omega_n$ [\unit{\radian\per\second}] & 1400 & $R_{sc,4}$ [\unit{\milli\ohm}] & 16.1 \\
        \hline
	\end{tabular}\vspace{-0.3cm}
    \end{center}
\end{table}

The FOC algorithm extended by the designed ISC mitigation was implemented on an Infineon AURIX TC397 application kit driving an EVAL-M1-IR2214 power stage at $T_s = 100$~\unit{\micro\second}. During the experiments, the PMSM drive operated in the speed mode, and the motor shaft was coupled through an NCTE 2200-17.5 torque sensor to a dynamometer, which imposed the prescribed load torque. The analog output of the torque sensor was synchronously acquired by the AURIX microcontroller and, together with the estimated speed, used to evaluate the mechanical power at the shaft $p_{shaft}$. The parameters of the PMSM, drive system, and FIU are summarized in Tab.~\ref{tab:params}, where $\omega_n$ describes the nominal electrical angular velocity of the motor.



To quantify the mitigation effect independently of the controller's internal estimates, the validation further employed power measurements. The objective was to verify whether the proposed mitigation reduces the motor input power $p_e$ while maintaining the same mechanical output during the fault and whether this global improvement is accompanied by a decrease in the local resistive loss of the shorted winding segment $p_{res}^{seg}$. We used a HIOKI PW8001 power analyzer equipped with a U7005 input unit and CT6872 current sensors to acquire the electrical quantities. The analyzer evaluated and logged $p_e$ over 50~\unit{\milli\second} integration windows using three channels, each pairing one phase current with the corresponding phase voltage measured with respect to the motor's accessible star point. To assess the local thermal effect of an ISC emulated in phase $a$, we complemented the acquired $i_a$ with additional FIU-related measurements. Specifically, the analyzer measured the FIU branch (fault) current $i_f$, the difference current $i_a-i_f$, and the FIU voltage $u_f$. The difference current was obtained by routing the phase $a$ conductor and the FIU branch conductor through the same current sensor in opposite directions. Using the RMS values evaluated over the same 50~\unit{\milli\second} integration window, we subsequently reconstructed the local resistive loss of the shorted segment, as in
\begin{equation}
	p_{res}^{seg}
	= (R_{seg}-R_{seg,f}) I_a^2
	+
	R_{seg,f} I_{a-f}^2
	+
	I_f U_f
	\label{eqn:pseg_res}
\end{equation}
where $I_a$, $I_f$, $I_{a-f}$, and $U_f$ are the RMS values of $i_a$, $i_f$, $i_a-i_f$, and $u_f$, respectively. The resistances $R_{seg} = 128$~\unit{\milli\ohm} and $R_{seg,f} = 40$~\unit{\milli\ohm} for $\sigma = 6/25$ then denote the total resistance of the tapped segment and the resistance of its shorted part. Both values were measured in advance using a HIOKI LCR HiTESTER 3511-50.

\begin{figure*}[t]
   	\begin{center}
   \includegraphics[]{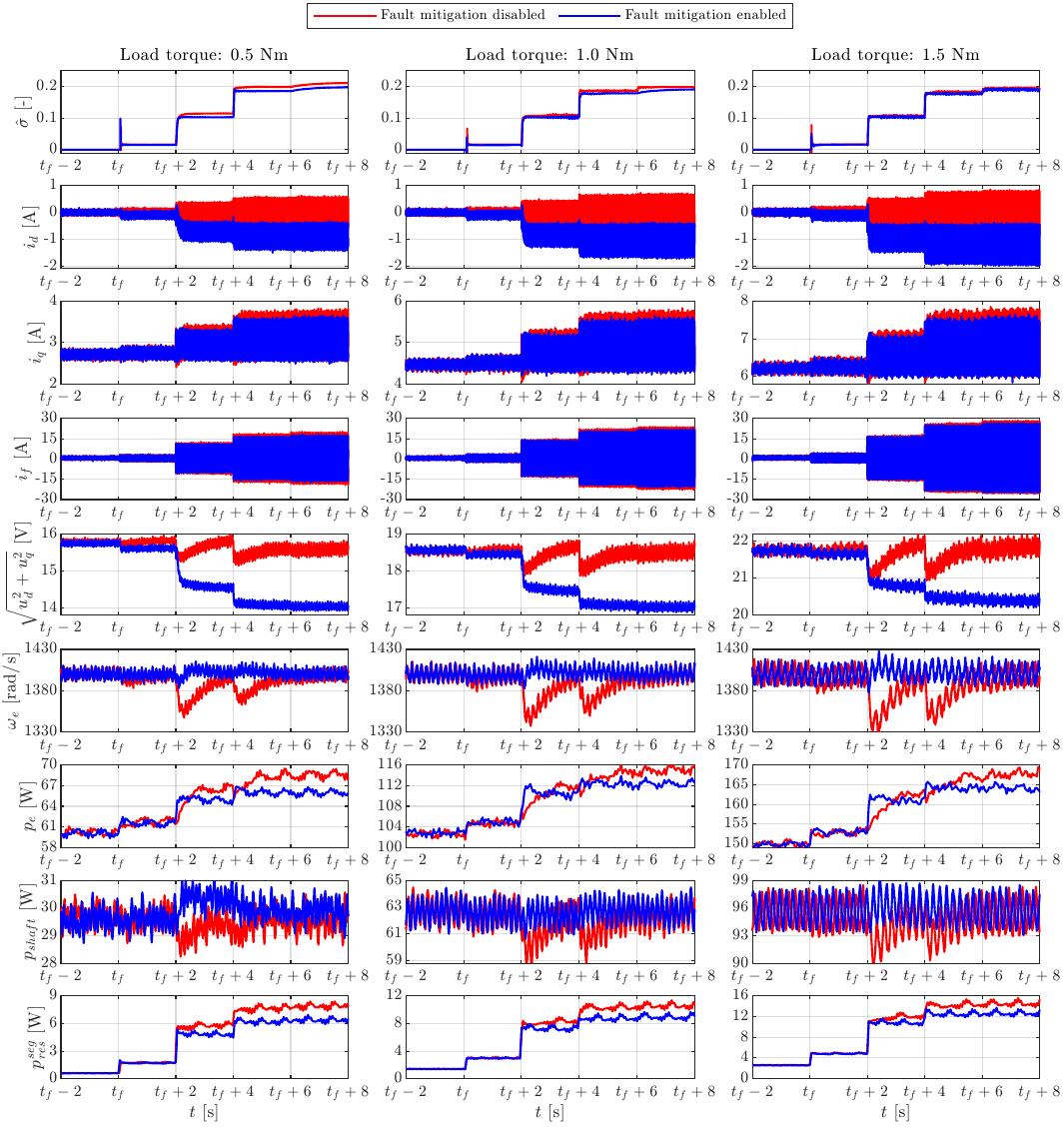}
   	\end{center}\vspace{-0.6cm}
   	\caption{The real-time FOC response to stepped $R_{sc}$ with disabled/enabled mitigation at $\omega_e=1400$~\unit{\radian\per\second}.}\vspace{-0.35cm}
   	\label{fig:mit1400}
   \end{figure*}
\begin{figure*}[t]
   	\begin{center}
   \includegraphics[]{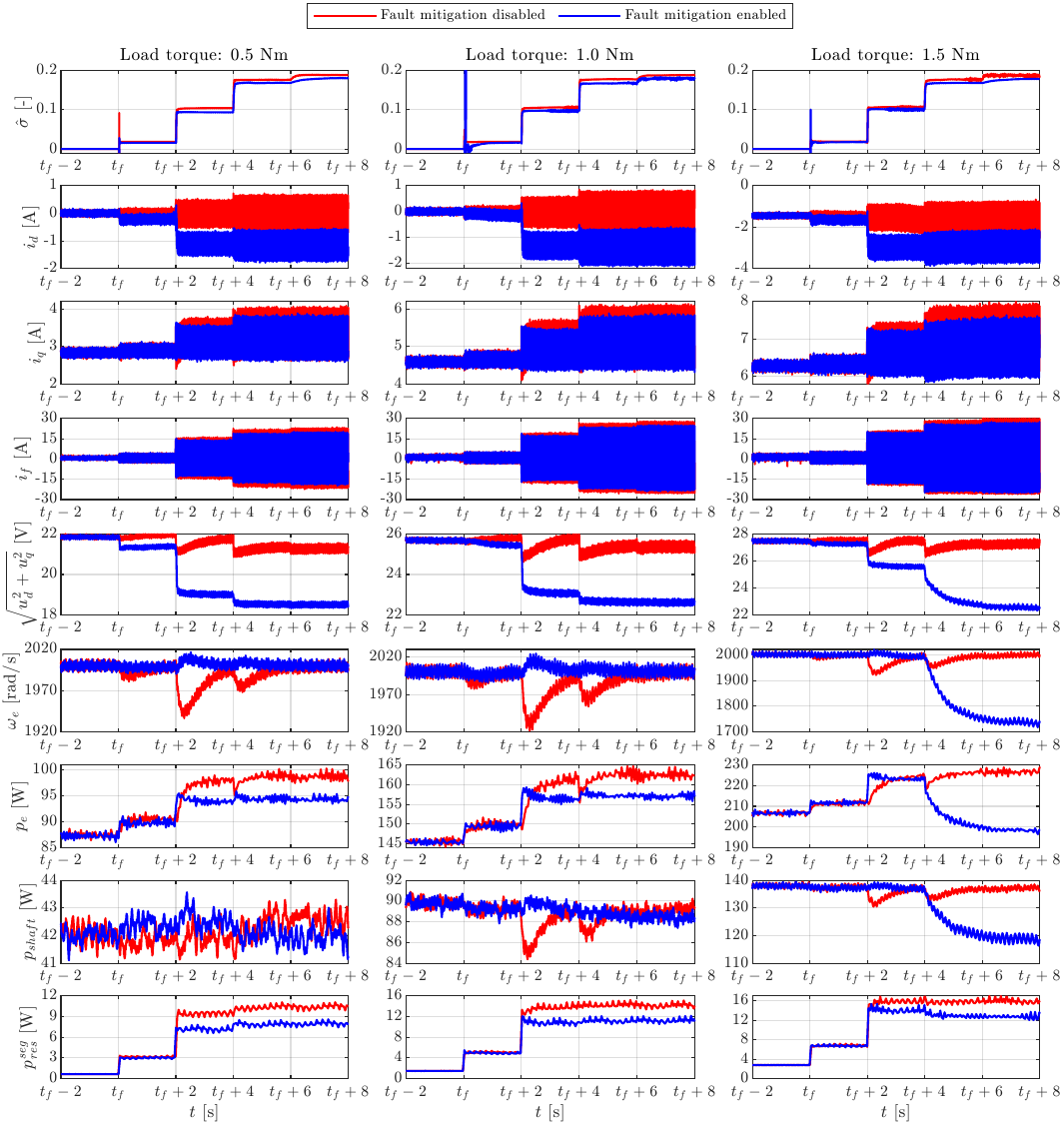}
   	\end{center}\vspace{-0.6cm}
   	\caption{The real-time FOC response to stepped $R_{sc}$ with disabled/enabled mitigation at $\omega_e=2000$~\unit{\radian\per\second}.}\vspace{-0.35cm}
   	\label{fig:mit2000}
   \end{figure*}

For each operating point, the validation compared two responses to the same emulated ISC: a baseline response with the mitigation disabled and a mitigated response with the proposed algorithm enabled. In both cases, the drive was operated at the same speed reference (1400 or 2000~\unit{\radian\per\second}) and the dynamometer imposed the same load torque (0.5, 1, or 1.5~\unit{\newton\meter}). The fault was introduced at time $t_f$ with $\sigma = 6/25$, and the FIU resistance was then stepped through $R_{\mathrm{sc},1}$--$R_{\mathrm{sc},4}$, with each value held for 2~\unit{\second}. Consequently, not only the method's loss-reduction capability but also its real-time adaptation to changing fault conditions were tested. The outcomes of the validation experiments are then depicted in Fig.~\ref{fig:mit1400} and Fig.~\ref{fig:mit2000}. The quantity $\hat{\sigma}$ represents the estimated fault severity, evaluated under the assumption of zero short circuit resistance \cite{DTM_fault_diag, Thesis}, i.e., $\hat{\sigma}=\sigma\left(1 - n_s R_{sc}/(n_p R_s \sigma + n_s R_{sc})\right)$.


As shown in Fig.~\ref{fig:mit1400} and Fig.~\ref{fig:mit2000}, with the mitigation disabled, each reduction in $R_{sc}$, captured by the increase of $\hat{\sigma}$, elevates the fault current, which distorts the total \textit{dq} currents. Although the current controller rapidly compensates for this distortion, the torque is produced only by the fault-free \textit{q}-axis component. Consequently, regulating the total $i_q$ effectively reduces the torque-producing term, and the missing torque is recovered only through the outer speed loop. This results in the observed speed drop after the fault steps. With the mitigation enabled, the current controller acts on the estimated fault-free current parts, so the torque-producing current is restored directly, and the speed transient is strongly suppressed. In addition, the smoother voltage magnitude indicates that the fault-induced ripple is no longer fed back into the voltage command. 

The power measurements then confirm the loss-reduction capability of the proposed mitigation. For the most severe fault step in the non-derated operating points, the fault-induced increase in $p_e$ is reduced by approximately 23--36\%, while the fault-induced increase in $p_{res}^{seg}$ is reduced by approximately 18--27\%. Thus, the lower input power is accompanied by a clear reduction of the local thermal loading at the shorted segment while preserving the same mechanical output. Importantly, at $\omega_e=2000$~\unit{\radian\per\second} and 1.5~\unit{\newton\meter}, the loss constraint makes the original operating point infeasible. The controller, therefore, derates the drive, demonstrating the intended fail-degraded behavior in which thermal protection is prioritized over torque production. Additional datasets for $\sigma=3/25$ and $\sigma=9/25$ are shared on Zenodo \cite{Datasets}.

\section{Conclusion}
\label{sec:conclusion}
\begin{table}[t]
	\centering
	\caption{Comparison of control-based ISC mitigation strategies}
	\label{tab:ftc_comparison}
	\definecolor{mygreen}{RGB}{50,205,50}
	\newcommand{\cmark}{\textcolor{mygreen}{\checkmark}}
	\newcommand{\xmark}{\textcolor{red}{$\times$}}
	\newcommand{\pmark}{$\triangle$}
	\setlength{\tabcolsep}{1.4pt}
	\renewcommand{\arraystretch}{1.08}
	\scriptsize
	
	\begin{tabular}{@{}>{\raggedright\arraybackslash}m{3.55cm}||
		*{8}{>{\centering\arraybackslash}m{0.45cm}}
		>{\centering\arraybackslash}m{0.60cm}@{}}
		\hline
		Feature 
		& \cite{PMSM_FD_FTC_Integration_TTE}
		& \cite{ITSC_Online_SOA_TEC}
		& \cite{ITSC_AdaptiveFTC_TIM}
		& \cite{ITF_MTPL_TIE}
		& \cite{DTC_SPMSM_ITSC_TorqueRipple_TPEL}
		& \cite{ITSC_MTPCC_TPEL}
		& \cite{CommonPredictiveModel_ITF_TTE}
		& \cite{DTPPMSM_SubspaceResiduals_TPEL}
		& Prop.\\
		\hline\hline
		
		Standard three-phase PMSM
		& \cmark & \cmark & \cmark & \cmark & \cmark & \cmark & \cmark & \xmark & \cmark\\
		
		No additional diagnostic hardware
		& \xmark & \cmark & \cmark & \pmark & \xmark & \pmark & \xmark & \cmark & \cmark\\
		
		No exact fault current required
		& \cmark & \xmark & \cmark & \pmark & \cmark & \xmark & \cmark & \cmark & \cmark\\
		
		Thermal/loss mitigation
		& \xmark & \cmark & \xmark & \cmark & \xmark & \cmark & \xmark & \xmark & \cmark\\
		
		Explicit resistive-loss constraint
		& \xmark & \xmark & \xmark & \cmark & \xmark & \xmark & \xmark & \xmark & \cmark\\
		
		Torque-ripple suppression
		& \cmark & \xmark & \cmark & \xmark & \cmark & \xmark & \cmark & \cmark & \cmark\\
		
		Diagnostics-control integration
		& \cmark & \pmark & \cmark & \pmark & \cmark & \xmark & \pmark & \cmark & \cmark\\
		
		Conventional FOC integration
		& \cmark & \xmark & \cmark & \cmark & \xmark & \cmark & \xmark & \pmark & \cmark\\
		
		Discrete-time formulation
		& \xmark & \pmark & \pmark & \xmark & \xmark & \xmark & \cmark & \xmark & \cmark\\
		
		Saliency considered: $L_d \neq L_q$
		& \xmark & \xmark & \pmark & \cmark & \xmark & \cmark & \cmark & \cmark & \xmark\\
		\hline
	\end{tabular}
	
	\vspace{0.5mm}
	\parbox{\columnwidth}{\scriptsize
	\cmark: addressed; 
	\pmark: partially addressed, assumed, or platform-dependent; 
	\xmark: not addressed.
	}
	\vspace{-0.3cm}
\end{table}
This article presented control-based ISC mitigation for PMSM drives. The method used diagnostic information available from the standard control loop in two complementary ways: to generate loss-constrained post-ISC current references and to provide fault-free current feedback for the current controller. As a result, the drive reduced fault-induced resistive losses while attenuating the torque disturbance. 

The method is compared with the existing control-based ISC mitigation strategies in Tab.~\ref{tab:ftc_comparison}, exposing that thermal-stress reduction and torque-ripple compensation are usually treated separately. The designed mitigation combines both these objectives in a conventional FOC structure without additional diagnostic hardware. Its main limitation is the equivalent-inductance assumption, which neglects saliency but remains consistent with the diagnostic stage, where velocity steady state operation usually does not provide sufficient excitation to separate $L_d$ and $L_q$ reliably.

\section*{Acknowledgment}
The MATLAB-compiled datasets attributed to this study are publicly available on Zenodo \cite{Datasets}.

\bibliographystyle{IEEEtranDOI}
\bibliography{references}

\vspace{-1.3cm}
\begin{IEEEbiography}[{\includegraphics[width=1in,height=1.25in,clip,keepaspectratio]{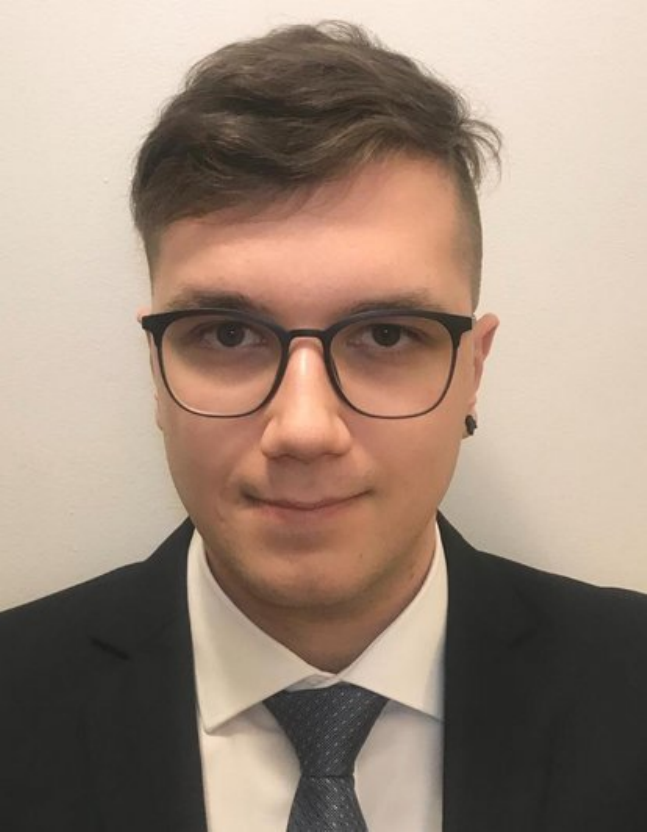}}]
{Lukas Zezula} was born in Brno, Czech Republic, in 1998. He received the M.Sc. degree in cybernetics, control and measurements; the M.Sc. degree in strategic company development; and the Ph.D. degree in cybernetics, control and measurements from Brno University of Technology, Brno, Czech Republic, in 2022, 2024, and 2025, respectively.

Currently, he is a Researcher with the Central European Institute of Technology, Brno University of Technology. His research interests include the continuous and discrete-time modeling of ac machines under faults, diagnostics in ac electric motors, and fault mitigation in electric drives.

Dr. Zezula received the IEEE Industrial Electronics Society Electric Machines Technical Committee Prize Paper Award at IECON 2024.
\end{IEEEbiography}

\vspace{-1.3cm}
\begin{IEEEbiography}[{\includegraphics[width=1in,height=1.25in,clip,keepaspectratio]{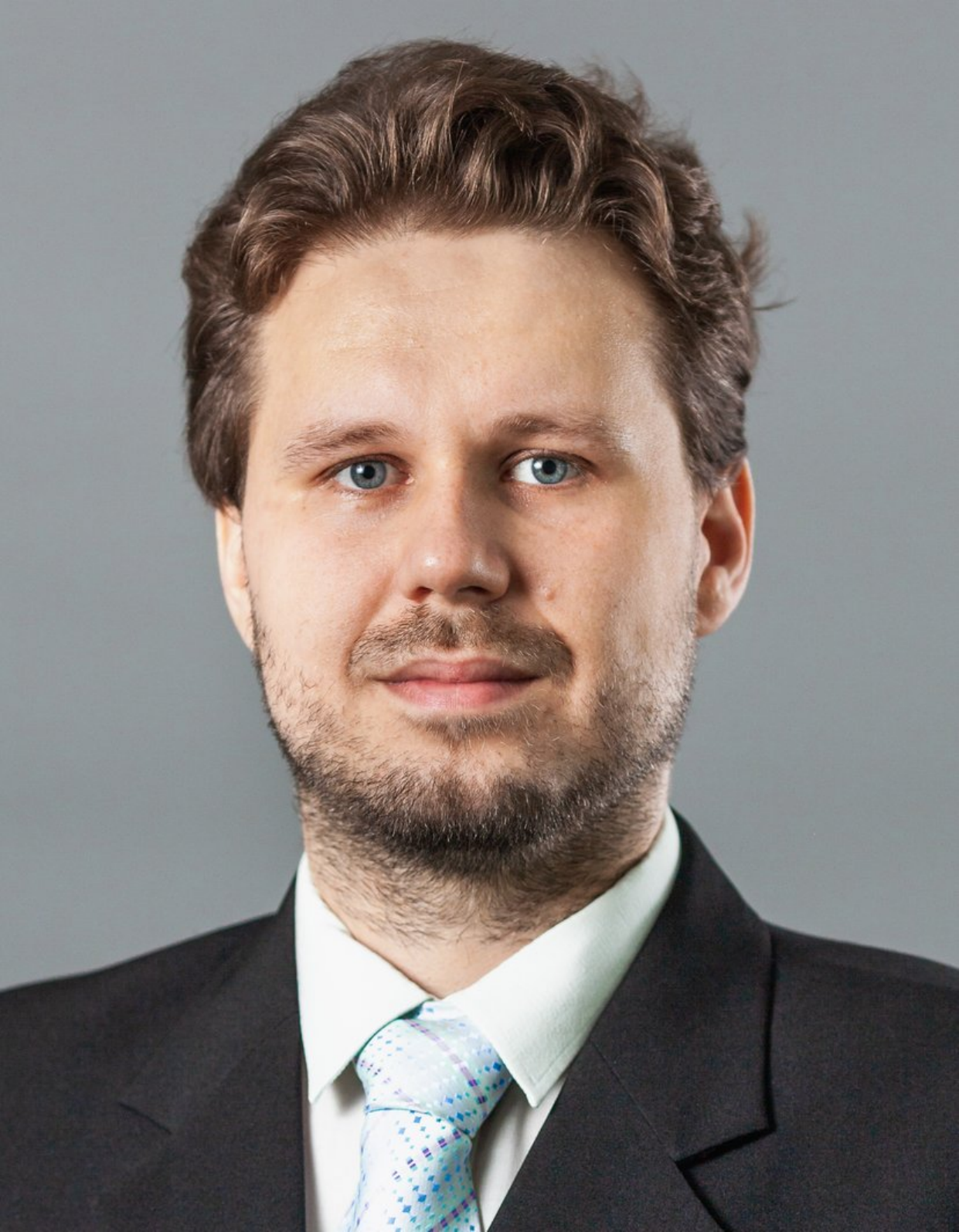}}]
{Matus Kozovsky} was born in Bratislava, Slovak Republic, in 1991. He received the Ph.D. degree in cybernetics, control and measurements from Brno University of Technology, Brno, Czech Republic, in 2021.

Currently, he is a Researcher of SW development with the Central European Institute of Technology, Brno University of Technology. His research interests include the fault-tolerant control of ac electric motors, high-speed control of electric motors, and inverter SW improvements.
\end{IEEEbiography}

\vspace{-1.3cm}
\begin{IEEEbiography}[{\includegraphics[width=1in,height=1.25in,clip,keepaspectratio]{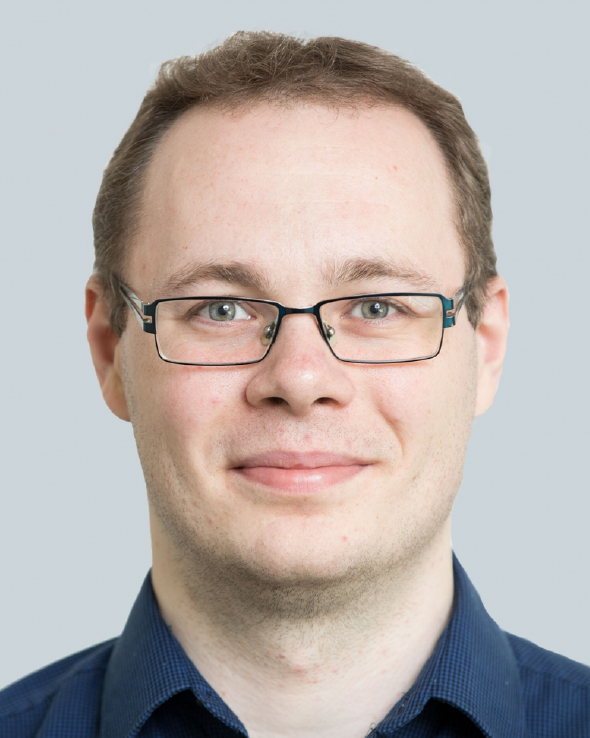}}]
{Ludek Buchta} was born in Brno, Czechoslovakia, in 1987. He received the Ph.D. degree in cybernetics, control and measurements from Brno University of Technology, Brno, Czech Republic, in 2019.
	
Currently, he is a Researcher with the Central European Institute of Technology, Brno University of Technology. His research interests include compensating the non-linearity of voltage source inverters, implementing neural network algorithms, and developing advanced algorithms to control ac electric motors.
\end{IEEEbiography}

\vspace{-1.3cm}
\begin{IEEEbiography}[{\includegraphics[width=1in,height=1.25in,clip,keepaspectratio]{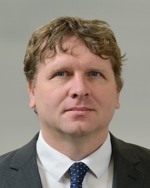}}]
{Petr Blaha} was born in Boskovice, Czechoslovakia, in 1973. He received the M.Sc. degree in cybernetics, control and measurements and a Ph.D. in cybernetics and informatics from Brno University of Technology, Brno, Czech Republic, in 1996 and 2001, respectively.

Currently, He is a Senior Researcher with the Central European Institute of Technology, Brno University of Technology, Brno, Czech Republic. Since 2007, he has been an Associate Professor with Brno University of Technology. His research interests include the modeling, diagnostics and advanced control of ac electric drives and the fault-tolerant control of electric motor drives.

Dr. Blaha received the IEEE Industrial Electronics Society Electric Machines Technical Committee Prize Paper Award at IECON 2024.
\end{IEEEbiography}
\end{document}